\documentclass{spro2012}
\usepackage{natbib}
\def\apj{Ap.J.}                 
\def\apjl{Ap.J.L}                
\def\ssr{Sp.Sci.Revs.}
\def\aap{Astron.Astrophys.}

\begin{document}

\title{The Impulsive Phase in Solar Flares: Recent Multi-wavelength Results and their Implications for Microwave Modeling and Observations}
\author{
   Lyndsay \textsc{Fletcher}\altaffilmark{1}
   and
   Paulo~J.~A. \textsc{Sim\~oes}\altaffilmark{1}
 }

 \altaffiltext{1}{SUPA School of Physics and Astronomy, University of Glasgow, Glasgow G12 8QQ, United Kingdom}

 \email{lyndsay.fletcher@glasgow.ac.uk}

\maketitle

\begin{abstract}
This short paper reviews several recent key observations of the processes occurring in the lower atmosphere (chromosphere and photosphere) during flares. These are: evidence for compact and fragmentary structure in the flare chromosphere,  the conditions in optical flare footpoints, step-like variations in the magnetic field during the flare impulsive phase, and hot, dense `chromospheric' footpoints. The implications of these observations for microwaves are also discussed.
\end{abstract}

\section{Introduction}
Via imaging, spectroscopy and time-series analysis, the microwave part of the spectrum provides vital information on the properties of flare-accelerated particles and the plasma and the magnetic field in which their emission is formed. Although there are considerable complexities in modeling and interpreting the data, microwaves are uniquely rich in diagnostic information and are crucial for flare studies. However, flares are characterised in part by the fact that - for the few minutes of the impulsive phase at least -  emission is generated across the entire electromagnetic spectrum. Therefore we have the ability to set our microwave observations in context, though in practice the number of flares with excellent multi-wavelength coverage including microwaves remains small. This highlights the ongoing need for continued operation of facilities such as the Nobeyama Radioheliograph (NoRH) and Radio Polarimeters (NoRP) in the current era of multi-wavelength observations. In this short paper we will review some multi-wavelength flare observations, focusing on  recent results relevant to the flare impulsive phase. These include hints of fine structuring in chromospheric footpoints, very compact footpoint sources, rapid changes in the tilt of the magnetic field during the flare impulsive phase, and hot chromospheric footpoints.  In the light of these results we will speculate on what the combination of multi-wavelength and microwave flare data can potentially bring. The `natural' partner for microwave emission is hard X-rays, and an extensive review of the strengths of combining radio and hard X-ray data can be found in \cite{2011SSRv..159..225W}.

\section{Structure of HXR footpoints}\label{sect:hxr}
\begin{figure}
\begin{center}
 \FigureFile(80mm,40mm){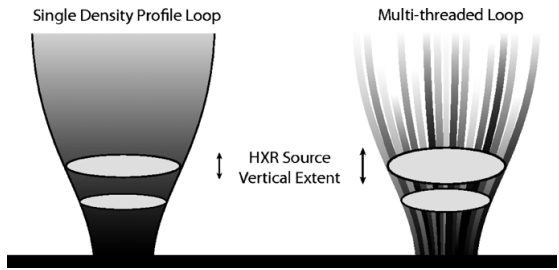}
  \end{center}
\caption{Illustration from Kontar et al. (2010) of how a multi-threaded loop, composed of numerous strands of different density scale-heights, can broaden a HXR source as observed, while preserving its centroid position.} \label{fig:strands}
\end{figure}

The most directly interpretable signature of non-thermal electrons in solar flares are the non-thermal hard X-rays (HXRs) emitted in bremsstrahlung interactions, particularly in the dense plasma of the solar chromosphere. HXR emission from the chromosphere is usually interpreted in terms of the collisional thick target model (CTTM), which means that the emission is generated as electrons slow down, under the influence of collisions only, and stop within the target. To interpret a given observation in this way also requires that the slowing-down timescale (fractions of a second) is less than the integration time used to make an image or spectrum, which is generally the case. Under the assumptions of the collisional thick target model, the total number of electrons that must be injected into the thick target to produce the observed spectrum can be inferred, in a way that does not depend on the precise density structure of the atmosphere. However, with imaging from the Ramaty High Energy Solar Spectroscopic Imager \citep[RHESSI, ][]{2002SoPh..210....3L} the density structure can be probed in more detail. The systematic offset of source position (first moment of the source intensity distribution) as a function of energy when interpreted in the CTTM gives a value for the target density scale-height \citep{2002SoPh..210..383A}. The source full-width at half-maximum intensity (2nd moment) as a function of energy can also be measured \citep{2011ApJ...735...42B}, and compared with the predictions of the CTTM; doing so is rather interesting as it is not possible to easily reconcile the modelled and observed behaviour of this quantity. The sources have a much larger observed FWHM than straightforward models of an electron distribution in a `monolithic' loop would predict \citep{2012ApJ...752....4B}. This may point to much finer sub-structure in the chromosphere composed of multiple sub-resolution structures of different density scale-heights, as shown in Figure~\ref{fig:strands}, such that the mean source position as a function of energy as expected from the CTTM is preserved, but the `variance' is increased \citep{2010ApJ...717..250K}. The implication is that chromospheric sub-structures on scales of a tenth of an arcsecond or less could exist. The fine-structuring would presumably have a coronal counterpart. However, the result may also point to problems with the applicability of the CTTM.  

Another important result to emerge from RHESSI is the inference, using the photosphere's albedo to HXR photons, that the electron angular distribution where the HXR footpoint emission is produced is consistent with one in which as many electrons are traveling `upwards' as `downwards'  \citep{2006ApJ...653L.149K}. Again this presents a challenge to the CTTM. The possibilities of developing such an electron distribution from a combination of scattering (collisional and non-collisional) and magnetic mirroring in the chromosphere are still being investigated, but already it is clear that  considerable fine-tuning of the electron, field and density parameters is needed to recover the observational results. 

The ongoing investigations into the causes of these observationally deduced properties of the electron spatial and energy distributions have yet to be concluded, but it is clear that a model of a downward-beamed `monolithic' electron beam entering a simple, uniform, collisional chromosphere, is not  correct. The electron distribution looks likely to be finely structured in space, and probably also time (though some average properties can be recovered), and may have a complex angular distribution. Co-ordinated observations with flares in the optical also make clear that electrons arrive at the chromosphere over small patches. It is not clear that optical footpoints are resolved on a scale of 1$"$ \citep{2006SoPh..234...79H}, so the typical optical footpoint is more like $10^{16}{\rm cm}^{2}$, rather than the $10^{18}{\rm cm}^{2}$ often used as a `canonical' footpoint size (see Section~\ref{sect:optical}). This may have significant implications for the usual `trap-plus-precipitation' models used in microwave modeling.

\section{The optical flare}~\label{sect:optical}
Optical, or `white light' flares, previously thought of primarily as accompanying large flares, are in fact common phenomena, but ill-understood. The flare optical and UV emission contains the majority of the flare's radiation output, and a wealth of spectral lines are available to probe the conditions in the flare chromosphere. For those reasons one would expect this part of the spectrum to have received more attention. However, optical flares are difficult to observe, as they have a low contrast compared to the bright photosphere, and spectroscopic observations require the good fortune to have a spectrometer slit on a flaring kernel at exactly the right time, which has in the past been rare. Nonetheless broad-band optical emission is observed in flares from C-class to X-class. The key is to have stable, high-cadence imaging or photometric observations which can be used to perform reliable differencing observations to pick up the faint flare signatures against the bright photospheric background. Doing this has revealed that optical footpoints are strongly correlated in space and time with HXR footpoints, and thus with the presence of large numbers of fast electrons \citep{2007ApJ...656.1187F} and that optical footpoints are very compact, with areas of around $10^{16}{\rm cm}^2$ or perhaps less. The energy contained in the white-light continuum is around 70\% of the total flare energy, independent of the flare class \citep{2011A&A...530A..84K}.

The emission mechanism of this broad-band optical radiation is not known. It seems unlikely that it is due to direct heating of the photosphere by electrons accelerated in the corona, as the typical electron energy required to reach the photosphere is around 2~MeV, assuming a column mass to the photosphere of 1~gm~cm$^{-2}$. An interesting recent analysis by \cite{2012ApJ...753L..26M} of a limb flare observed by RHESSI and also by one of the STEREO spacecraft uses careful triangulation to place both the flare optical emission and the HXR emission at 30-50~keV at only 300~km above the photosphere. This corresponds to an electron stopping energy of around 1~MeV. (This single observation has yet to be repeated). For the optical luminosity to be produced at or near the photosphere by electrons arriving in a beam from the corona, a large fraction of the electron energy in the injected spectrum would have to be above $\sim$~1~MeV. This is inconsistent with parameters for the electron distribution derived from HXR measurements. Microwave and millimeter observations are informative here, as the emission is generated by electrons in the 100~keV-plus energy range, and it is interesting that these observations suggest that the spectrum may be substantially harder at energies above a few 100~keV \citep[e.g.][]{1994ApJS...90..599K} than would be implied by the continuation of the HXR-emitting spectrum. However, this is still not adequate. For example, application of gyrosynchrotron models to microwave emission by \cite{2003ApJ...595L.111W} for the large flare SOL2002-07-23T00:35 (X4.8) results in a non-thermal electron density distribution of $n_e = 5\times10^6 (B/200 G)^{-2.1}(E/20{\rm keV})^{-2.5}{\rm electrons~cm^{-3}keV^{-1}}$, corresponding to an electron energy density above $E_c =$1~MeV of $ \int_{E_c}^\infty E n_e  dE \sim 1~{\rm erg~cm}^{-3}$ if we assume $B = 200$G. This is far too small compared to the energy density of the photospheric plasma (around $10^4~{\rm erg~cm}^{-3}$) to produce an observable optical intensity perturbation. For now the observation of \cite{2012ApJ...753L..26M} remains a puzzle.

Another mechanism for producing optical emission is the free-bound continuum that results from the ionisation and recombination in a flare-heated chromosphere (note, the heating can be, but does not need to be, provided by non-thermal electrons).  The UV (Balmer and Lyman) continua may also penetrate downwards and backwarm the photosphere, effectively recycling this radiation as optical emission. By indirect means, optically thin emission has been deduced in one flare with an extended white-light ribbon \citep{2010ApJ...722.1514P}. This would be consistent with free-bound emission from a hot plasma. The temperature of the emitting plasma should be high enough that the hydrogen is substantially ionised, i.e. above $\sim 10,000 - 15,000$K. The brightness temperature of the surrounding non-flaring chromosphere is also in this range, so associated microwave emission would not be visible unless the free-bound emitting plasma were hotter. Moreover,  in the free-bound model the electron density in the region emitting in the optical may be around $10^{13}\rm{cm}^{-3}$ \citep{2003ApJ...595..483M}, so any microwaves generated here with frequencies below the corresponding plasma frequency ($\sim$ 28GHz) will not propagate. 

Although we do not expect to make direct microwave observations of the plasma that radiates the optical emission, the observed optical source properties, coupled with those inferred from hard X-ray spectra, have implications for beam parameters which should be recognised in future microwave modeling. The optical footpoint area is small, and in the event presented by \cite{2011ApJ...739...96K} it is consistent with an unresolved HXR footpoint size. Interpreted in the CTTM the {\emph{non-thermal}} electron beam density at the location of HXR emission in this event is at least $10^{10}$ electrons~${\rm cm}^{-3}$ above 20~keV. This is consistent with the value inferred in another X-flare by \cite{2003ApJ...595L.111W}  from the microwave emission in another large flare, but this time in the corona, where the 17~GHz emission is located. If magnetic mirroring occurs, due to field convergence between the corona and chromosphere (as one might expect given the increasing evidence for very small footpoints) then the HXR-generating electrons in the chromosphere would represent only the component that can escape the magnetic trap, giving a lower limit to the overall non-thermal coronal electron density in the coronal loop. On the other hand if the electrons were beamed exactly along the magnetic field then there would be no trapped component and, in a field that diverges into the corona, the coronal beam density requirement would be reduced. But a highly-beamed distribution is at present inconsistent with the angular distributions inferred using photospheric HXR albedo (Section~\ref{sect:hxr}.)

Such high beam densities challenge electron transport models, but may have some bearing on electron spectra relevant to microwave and X-ray comparisons, because beam-return current instabilities can substantially distort the electron spectrum. Unless the background plasma through which the beam passes is much denser than the beam, such that the return current speed is low, the beam-return current system will be subject to plasma instabilities causing the majority of its energy to be dissipated as electron and ion heating, via wave generation.  Analytical considerations suggest that a beam with even a fraction of the density implied by the combination of X-ray and optical observations should, together with its return current system become unstable to the ion-acoustic instability (when its return-current speed is greater than the ion-acoustic speed), with the beam energy redistributed in heating, unless the ambient density is much larger than the beam density \citep{1976SoPh...48..197H}. Numerical simulations of the beam-return current system are now very elaborate, including Vlasov and PIC simulations, in magnetised and non-magnetised scenarios. For example, work by \cite{2011PhPl...18b2308L} and \cite{2012A&A...544A.148K} indicates that the majority of the beam energy  - around 70\% - is redistributed as plasma heating, but that the remaining fraction may be available to re-accelerate higher energy electrons. This is interesting for the comparison between HXR and microwave radiation, since it suggests that the lower energy X-ray generating electrons and the higher energy microwave-emitting electrons need not be described by one spectral index. A number of joint studies of HXR and microwave flare conclude that there is a substantial difference in the electron spectral indices at low energy and high energy  \citep[e.g.][]{1991SoPh..132..125N,1994ApJS...90..599K,2003ApJ...595L.111W}, where microwave emission interpreted as optically-thin gyrosynchrotron radiation implies that electron spectra at high energies (above $\sim 200$keV) are substantially flatter than at low energies.

\section{Impulsive-phase variations in the line-of-sight magnetic field}

\begin{figure}
\begin{center}
 \FigureFile(80mm,80mm){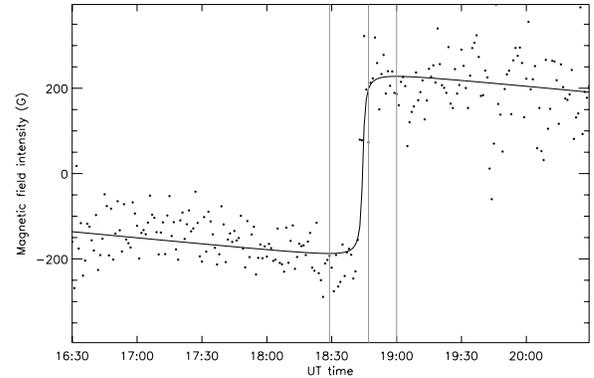}
  \end{center}
\caption{Measurements of the line-of-sight magnetic field with GONG at a single representative pixel location in the flare SOL2006-12-06T18:45 (X6.5), showing a step-like non-reversing change. The vertical lines mark the GOES X-ray start, peak and end times. From \cite{2010ApJ...724.1218P}} \label{fig:petriesudol}
\end{figure}

The impulsive phase of a solar flare has detectable consequences for the low solar atmosphere, i.e. the photosphere, apart from the possible photospheric origin for optical flares. It is now well established that significant abrupt (step-like), non-reversing change in the line-of-sight photospheric field occurs for major (X-class and M-class) flares, co-temporal with the flare impulsive phase \citep[e.g.][]{2005ApJ...635..647S,2012ApJ...756..144C}. An example of such a change is shown in Figure~\ref{fig:petriesudol}. The location of this change can be in the umbra, the penumbra or elsewhere in the active region, and  is co-spatial with increases in UV footpoint intensity \citep{2010ApJ...724.1218P,2012ApJ...760...29J}.  The onset of the GOES soft X-ray emission leads the field changes \citep{2012ApJ...756..144C} as do peaks in the UV intensity - by on average 4 minutes \citep{2012ApJ...760...29J}. As remarked on by these authors, the timing pattern is consistent with the flare causing the photospheric field changes, and not vice versa. The HXR footpoints are associated with some, but not all, locations \citep[e.g.][]{2008SoPh..251..613M,2011ApJ...739...71M}, though no systematic study of this has yet been carried out. The change of the line-of-sight field is taken as a sign that the coronal magnetic field is reconfigured by the flare, as shear relaxes or the field shrinks, and that this reconfiguration propagates through the atmosphere to the field's anchor points in the photosphere. The tendency is for the field to become more horizontal \citep{2010ApJ...724.1218P}. 

\begin{figure}
\begin{center}
 \FigureFile(80mm,80mm){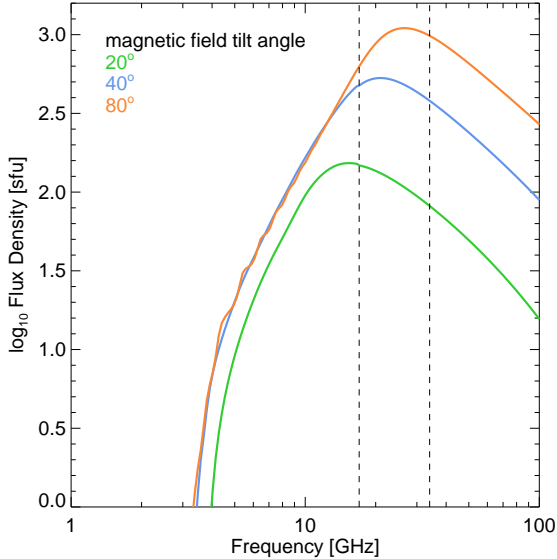}
  \end{center}
\caption{Calculated microwave spectrum from the an electron distribution in conditions appropriate for the lower corona or heated footpoints, under three conditions of viewing angle to the line of sight (tilt angle). The curves correspond to tilt angle 80$^\circ$ (upper), 40$^\circ$ (upper) and 20$^\circ$ (upper)} \label{fig:tilt}
\end{figure}

The change in the field direction over the duration of the flare impulsive phase corresponds of course to a variation in the viewing angle. Since gyrosynchrotron emission from the high-energy non-thermal electrons which are also present during the flare impulsive phase is anisotropic this variation in the viewing angle will influence the microwave emission observed.  We do not know of any comparisons yet being made between field changes and variations in the microwave, and indeed it might be difficult to disentangle variations due to changes in the magnetic field direction from those due to variations in the intrinsic properties of the population of emitting electrons. However, we can anticipate the effects. In Figure~\ref{fig:tilt} we show the variation in the microwave spectrum as the viewing angle changes while all other parameters of the source stay the same. Vertical lines on this plot show the NoRH observing frequencies. The emission is calculated for a non-thermal electron density of $10^8{\rm cm}^{-3}$ and electron spectral index of 3 in a compact source of diameter 5", thickness 5000~km, and temperature 5MK in a magnetic field of 500~G and ambient density of $10^{11}{\rm cm}^{-3}$. This could correspond to a gyrosynchrotron source near the footpoints of a coronal loop, or flare-heated upper chromosphere (see Section \ref{sect:hotfp}). The three curves correspond to the same field strength, viewed at an angle to the field direction of 20$^\circ$, 40$^\circ$ and 80$^\circ$. A decrease in the line-of-sight component of the field strength caused by an increasing field tilt leads to higher microwave intensity and an increase in the peak frequency. It would be interesting to search for a systematic effect such as this in the data, but in any such effort the many other parameters affecting the microwave spectrum must also be accounted for.

\section{Hot footpoints}~\label{sect:hotfp}

\begin{figure}
\begin{center}
 \FigureFile(90mm,80mm){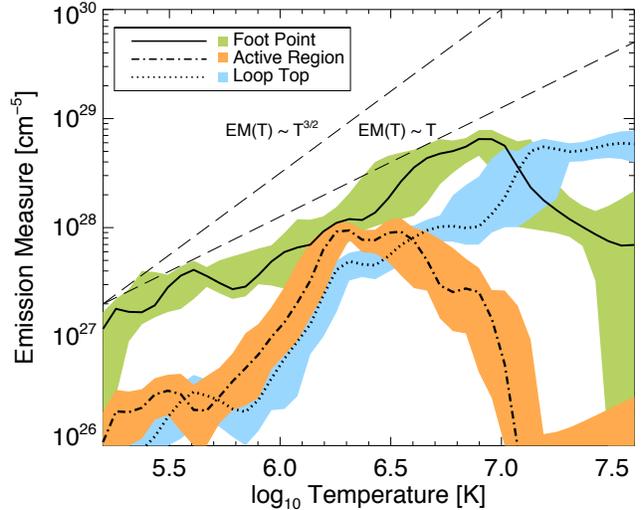}
  \end{center}
\caption{The emission measure distributions determined from {\it{Hinode}}/EIS observations of a flare footpoint (solid line), a flare looptop (dotted line), and average active region (dot-dashed line). The shaded areas gives the confidence limits of the EM reconstruction. The straight dashed lines show gradients 3/2 and 1. From \cite{2013arXiv1302.2514G}} \label{fig:fpems}
\end{figure}

The chromosphere in solar flares is strongly heated. This is readily seen in e.g. EUV images of flare ribbons which indicate plasma of at least a million degrees. However, it has been known at least since the days of {\it{Yohkoh}}, though not widely appreciated, that more extreme plasmas exist in the chromospheric footpoints during the flare impulsive phase. Impulsive soft X-ray footpoints observed via their bremsstrahlung emission by \cite{1994ApJ...422L..25H} and, in a large sample by \cite{2004A&A...415..377M} show temperatures close to 10~MK, and densities of at least a few times $10^{10}{\rm cm}^{-3}$ (depending on assumptions about their size). Using EUV spectroscopy of flare footpoints from {\it{Hinode}}/EIS \citep{2013arXiv1302.2514G} have determined the emission measure distribution for impulsive phase footpoints in a number of small (B- and C-class) events, and these typically also peak at 10~MK. An example of a footpoint emission measure distribution for a C1.1 flare is shown in Figure~\ref{fig:fpems}, compared with the loop emission measure distribution from the same time in this event, and non-flaring active region emission measure distribution. Independently, density diagnostics of this event return values of around $10^{11}\rm{cm}^{-3}$ at a temperature of 1.8~MK \citep{2011ApJ...740...70M}. Direct density diagnostics for higher temperatures were not available. The gradient of $\log EM - \log T$ for the footpoint is 1 in all cases studied; we note that this is consistent with conductive heating balanced by radiative cooling \citep{1973SoPh...33..341S}.

\begin{figure}
\begin{center}
 \FigureFile(80mm,80mm){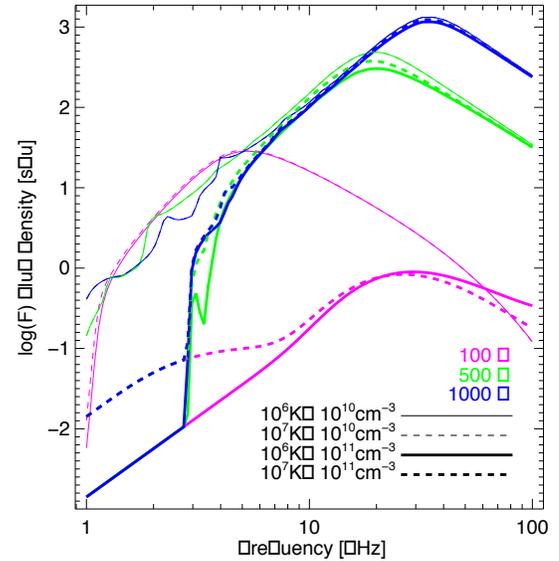}
  \end{center}
\caption{Calculated gyrosynchrotron spectrum from a non-thermal electron spectrum in a hot, thermal, compact footpoint plasma. Field strengths used are 1000~G (upper set of curves), 500~G (middle set) and 100~G (lower set). Solid lines correspond to a plasma temperature of $10^6$~K and dashed lines to $10^7$K. Thin lines correspond to the lower density of $10^{10}{\rm cm}^{-3}$ and thick lines to $10^{11}{\rm cm}^{-3}$.} \label{fig:hotfps}
\end{figure}

The consequences of these hot, dense compact footpoints for microwave footpoint emission have not been explored. In the usual microwave flare modelling, the characteristics of the {\emph{coronal}} flaring source are carefully studied, e.g. the inhomogeneities of the coronal field \citep[e.g.][]{2006A&A...453..729S} or the effects of pitch-angle distribution of the electrons on the emission \citep[e.g.][]{2010SoPh..266..109S}. But the modelling assumes that footpoints are rather cool as well as dense, and microwave emission will thus be free-free absorbed. The hot footpoint plasmas that we find are essentially at `coronal' temperatures but located at chromospheric altitudes and with density and magnetic field strength higher than typically found in the corona. They would be expected to produce intense compact sources, with spectral properties similar to those computed for coronal loops, and dominating any coronal emission in their optically-thin ranges due to higher densities and fields. The high plasma densities might however lead to Razin suppression at low frequencies. In Figure~\ref{fig:hotfps} we show calculations of footpoint emission for different temperatures and densities in the ranges deduced from the EUV observations. Of course, the observations also suggest that the hot footpoint plasma would be very inhomogeneous, with temperature varying by a factor 10 over a distance of probably 1000~km or so. So these spectra are for the moment only indicative. The parameters used in these calculations are: field strength of 100, 500 or 1000~G  (magenta, green and blue curves, respectively) and viewing angle of 45$^\circ$, isotropic electron distribution with flux spectral index $\delta = 3.6$, having a minimum electron energy of 200~keV and a maximum energy of 5~MeV. The source angular diameter is 5$"$ and depth along the line of sight is 2$"$, comparable to the depth of the chromosphere. The non-thermal electron density in the footpoint is around one part in $10^4$ of the background thermal density, or $6.9\times 10^6\rm{cm^{-3}}$. 

For the temperatures $10^6$K and $10^7$K (shown in Figure~\ref{fig:hotfps} by continuous and dashed curves) and densities  $10^{10}\rm{cm}^{-3}$ and $10^{11}\rm{cm}^{-3}$ (shown by thin and thick  lines) found from EUV and soft X-rays, observable footpoint microwave sources in the NoRP frequency ranges are predicted. The emission is mostly non-thermal gyrosynchrotron, and source intensity is determined mainly by the magnetic field strength, where stronger fields shift the spectrum peak towards higher frequencies with stronger emission \citep[e.g.][]{1989SoPh..120..351S}.  The contribution of the non-thermals to the microwave spectrum as shown in Figure~\ref{fig:hotfps} is greatly in excess of the thermal gyrosynchrotron which would be expected from these same plasma parameters. For a given field strength the intensity in the NoRH range, and NoRP above 3.8 GHz, is rather insensitive to the different values of density and temperature used here. The exception is in the weak-field case where the emission is affected by free-free absorption and Razin suppression. The Razin suppression \citep[e.g.][]{1969ApJ...158..753R} is significant for microwave frequencies below the Razin frequency $\nu_R=2\nu_p^2/(3\nu_B \sin \theta)$, which is below 5 GHz for all  cases considered, except for $B=100$G and $n_p=10^{11}\rm{cm}^{-3}$, where $\nu_R \approx 27$ GHz. The resulting spectra are thus  strongly suppressed, and show slighly different contributions from that produced by the free-free mechanism, for $T=10^6$K or $10^7$K. The steep emission decrease towards lower frequencies is caused by absorption below the plasma frequency (around 1 GHz and 3 GHz for densities of $10^{10}\rm{cm}^{-3}$ and $10^{11}\rm{cm}^{-3}$, respectively).

Imaging spectropic analysis of future observations with E-OVSA, coupled with EUV, UV and optical observations, will provide an excellent diagnostic tool for deriving the magnetic field, plasma  density and temperature in chromospheric flaring regions.
We should also note that the recently discovered sub-THz spectral component above 100 GHz \citep{2004ApJ...603L.121K} is likely to be  generated in the chromosphere, when considering the proposed radiation mechanisms \citep{2010ApJ...709L.127F}. In terms of gyrosynchrotron emission, it can be shown that
the second spectral peak can be produced in hot and dense footpoints with strong magnetic fields and strong Razin suppression, self-consistently with the typical microwave spectrum generated in the coronal source \citep{2013MCS_inprep}.

\section{Conclusions}
Observational understanding of the energetically dominant processes in the flare lower atmosphere during the impulsive phase, drawing on the many space- and ground-based instruments currently observing the Sun, is developing rapidly and in some unexpected directions. The new results presented here on the plasma and field parameters in the chromosphere during the flare impulsive phase are important for future microwave modelling, and the multi-wavelength data that we now have at our disposal must also be confronted with ongoing microwave imaging and spectra observations, which provides unique diagnostics of both plasma and field. To this end, the continued operation of the Nobeyama Radioheliograph and Radio Polarimeters remain crucial for our exploration and understanding of flares.

\section{Acknowledgements}
L.F. is very grateful to the conference organisers for the financial support which allowed her participation in this most simulating meeting. This work was supported by STFC grant ST/I001808/1 and by EC-funded FP7 project HESPE (FP7-2010-SPACE-1-263086).


\begin{thebibliography}{36}
\expandafter\ifx\csname natexlab\endcsname\relax\def\natexlab#1{#1}\fi
\expandafter\ifx\csname url\endcsname\relax
  \def\url#1{{\tt #1}}\fi

\bibitem[{Aschwanden} et~al.(2002){Aschwanden}, {Brown}, and
  {Kontar}]{2002SoPh..210..383A}
M.~J. {Aschwanden}, J.~C. {Brown}, and E.~P. {Kontar}.
\newblock {Chromospheric Height and Density Measurements in a Solar Flare
  Observed with RHESSI II. Data Analysis}.
\newblock {\em \solphys}, 210:\penalty0 383--405, November 2002.

\bibitem[{Battaglia} and {Kontar}(2011)]{2011ApJ...735...42B}
M.~{Battaglia} and E.~P. {Kontar}.
\newblock {Hard X-Ray Footpoint Sizes and Positions as Diagnostics of Flare
  Accelerated Energetic Electrons in the Low Solar Atmosphere}.
\newblock {\em \apj}, 735:\penalty0 42, July 2011.

\bibitem[{Battaglia} et~al.(2012){Battaglia}, {Kontar}, {Fletcher}, and
  {MacKinnon}]{2012ApJ...752....4B}
M.~{Battaglia}, E.~P. {Kontar}, L.~{Fletcher}, and A.~L. {MacKinnon}.
\newblock {Numerical Simulations of Chromospheric Hard X-Ray Source Sizes in
  Solar Flares}.
\newblock {\em \apj}, 752:\penalty0 4, June 2012.

\bibitem[{Cliver} et~al.(2012){Cliver}, {Petrie}, and
  {Ling}]{2012ApJ...756..144C}
E.~W. {Cliver}, G.~J.~D. {Petrie}, and A.~G. {Ling}.
\newblock {Abrupt Changes of the Photospheric Magnetic Field in Active Regions
  and the Impulsive Phase of Solar Flares}.
\newblock {\em \apj}, 756:\penalty0 144, September 2012.

\bibitem[{Fleishman} and {Kontar}(2010)]{2010ApJ...709L.127F}
G.~D. {Fleishman} and E.~P. {Kontar}.
\newblock {Sub-Thz Radiation Mechanisms in Solar Flares}.
\newblock {\em \apjl}, 709:\penalty0 L127--L132, February 2010.

\bibitem[{Fletcher} et~al.(2007){Fletcher}, {Hannah}, {Hudson}, and
  {Metcalf}]{2007ApJ...656.1187F}
L.~{Fletcher}, I.~G. {Hannah}, H.~S. {Hudson}, and T.~R. {Metcalf}.
\newblock {A TRACE White Light and RHESSI Hard X-Ray Study of Flare
  Energetics}.
\newblock {\em \apj}, 656:\penalty0 1187--1196, February 2007.

\bibitem[{Graham} et~al.(2013){Graham}, {Hannah}, {Fletcher}, and
  {Milligan}]{2013arXiv1302.2514G}
D.~R. {Graham}, I.~G. {Hannah}, L.~{Fletcher}, and R.~O. {Milligan}.
\newblock {The emission measure distribution of impulsive phase flare
  footpoints}.
\newblock {\em ArXiv e-prints}, February 2013.

\bibitem[{Hoyng} et~al.(1976){Hoyng}, {Brown}, and {van
  Beek}]{1976SoPh...48..197H}
P.~{Hoyng}, J.~C. {Brown}, and H.~F. {van Beek}.
\newblock {High time resolution analysis of solar hard X-ray flares observed on
  board the ESRO TD-1A satellite}.
\newblock {\em \solphys}, 48:\penalty0 197--254, June 1976.

\bibitem[{Hudson} et~al.(1994){Hudson}, {Strong}, {Dennis}, {Zarro}, {Inda},
  {Kosugi}, and {Sakao}]{1994ApJ...422L..25H}
H.~S. {Hudson}, K.~T. {Strong}, B.~R. {Dennis}, D.~{Zarro}, M.~{Inda},
  T.~{Kosugi}, and T.~{Sakao}.
\newblock {Impulsive behavior in solar soft X-radiation}.
\newblock {\em \apjl}, 422:\penalty0 L25--L27, February 1994.

\bibitem[{Johnstone} et~al.(2012){Johnstone}, {Petrie}, and
  {Sudol}]{2012ApJ...760...29J}
B.~M. {Johnstone}, G.~J.~D. {Petrie}, and J.~J. {Sudol}.
\newblock {Abrupt Longitudinal Magnetic Field Changes and Ultraviolet Emissions
  Accompanying Solar Flares}.
\newblock {\em \apj}, 760:\penalty0 29, November 2012.

\bibitem[{Karlick{\'y}} and {Kontar}(2012)]{2012A&A...544A.148K}
M.~{Karlick{\'y}} and E.~P. {Kontar}.
\newblock {Electron acceleration during three-dimensional relaxation of an
  electron beam-return current plasma system in a magnetic field}.
\newblock {\em \aap}, 544:\penalty0 A148, August 2012.

\bibitem[{Kaufmann} et~al.(2004){Kaufmann}, {Raulin}, {de Castro}, {Levato},
  {Gary}, {Costa}, {Marun}, {Pereyra}, {Silva}, and
  {Correia}]{2004ApJ...603L.121K}
P.~{Kaufmann}, J.-P. {Raulin}, C.~G.~G. {de Castro}, H.~{Levato}, D.~E. {Gary},
  J.~E.~R. {Costa}, A.~{Marun}, P.~{Pereyra}, A.~V.~R. {Silva}, and
  E.~{Correia}.
\newblock {A New Solar Burst Spectral Component Emitting Only in the Terahertz
  Range}.
\newblock {\em \apjl}, 603:\penalty0 L121--L124, March 2004.

\bibitem[{Kontar} and {Brown}(2006)]{2006ApJ...653L.149K}
E.~P. {Kontar} and J.~C. {Brown}.
\newblock {Stereoscopic Electron Spectroscopy of Solar Hard X-Ray Flares with a
  Single Spacecraft}.
\newblock {\em \apjl}, 653:\penalty0 L149--L152, December 2006.

\bibitem[{Kontar} et~al.(2010){Kontar}, {Hannah}, {Jeffrey}, and
  {Battaglia}]{2010ApJ...717..250K}
E.~P. {Kontar}, I.~G. {Hannah}, N.~L.~S. {Jeffrey}, and M.~{Battaglia}.
\newblock {The Sub-arcsecond Hard X-ray Structure of Loop Footpoints in a Solar
  Flare}.
\newblock {\em \apj}, 717:\penalty0 250--256, July 2010.

\bibitem[{Kretzschmar}(2011)]{2011A&A...530A..84K}
M.~{Kretzschmar}.
\newblock {The Sun as a star: observations of white-light flares}.
\newblock {\em \aap}, 530:\penalty0 A84, June 2011.

\bibitem[{Krucker} et~al.(2011){Krucker}, {Hudson}, {Jeffrey}, {Battaglia},
  {Kontar}, {Benz}, {Csillaghy}, and {Lin}]{2011ApJ...739...96K}
S.~{Krucker}, H.~S. {Hudson}, N.~L.~S. {Jeffrey}, M.~{Battaglia}, E.~P.
  {Kontar}, A.~O. {Benz}, A.~{Csillaghy}, and R.~P. {Lin}.
\newblock {High-resolution Imaging of Solar Flare Ribbons and Its Implication
  on the Thick-target Beam Model}.
\newblock {\em \apj}, 739:\penalty0 96, October 2011.

\bibitem[{Kundu} et~al.(1994){Kundu}, {White}, {Gopalswamy}, and
  {Lim}]{1994ApJS...90..599K}
M.~R. {Kundu}, S.~M. {White}, N.~{Gopalswamy}, and J.~{Lim}.
\newblock {Millimeter, microwave, hard X-ray, and soft X-ray observations of
  energetic electron populations in solar flares}.
\newblock {\em \apjs}, 90:\penalty0 599--610, February 1994.

\bibitem[{Lee} and {B{\"u}chner}(2011)]{2011PhPl...18b2308L}
K.~W. {Lee} and J.~{B{\"u}chner}.
\newblock {Turbulent anomalous transport and anisotropic electron heating in a
  return current system}.
\newblock {\em Physics of Plasmas}, 18\penalty0 (2):\penalty0 022308, February
  2011.

\bibitem[{Lin} et~al.(2002){Lin}, {Dennis}, {Hurford}, {Smith}, {Zehnder},
  {Harvey}, {Curtis}, {Pankow}, {Turin}, {Bester}, {Csillaghy}, {Lewis},
  {Madden}, {van Beek}, {Appleby}, {Raudorf}, {McTiernan}, {Ramaty}, {Schmahl},
  {Schwartz}, {Krucker}, {Abiad}, {Quinn}, {Berg}, {Hashii}, {Sterling},
  {Jackson}, {Pratt}, {Campbell}, {Malone}, {Landis}, {Barrington-Leigh},
  {Slassi-Sennou}, {Cork}, {Clark}, {Amato}, {Orwig}, {Boyle}, {Banks},
  {Shirey}, {Tolbert}, {Zarro}, {Snow}, {Thomsen}, {Henneck}, {McHedlishvili},
  {Ming}, {Fivian}, {Jordan}, {Wanner}, {Crubb}, {Preble}, {Matranga}, {Benz},
  {Hudson}, {Canfield}, {Holman}, {Crannell}, {Kosugi}, {Emslie}, {Vilmer},
  {Brown}, {Johns-Krull}, {Aschwanden}, {Metcalf}, and
  {Conway}]{2002SoPh..210....3L}
R.~P. {Lin}, B.~R. {Dennis}, G.~J. {Hurford}, D.~M. {Smith}, A.~{Zehnder},
  P.~R. {Harvey}, D.~W. {Curtis}, D.~{Pankow}, P.~{Turin}, M.~{Bester},
  A.~{Csillaghy}, M.~{Lewis}, N.~{Madden}, H.~F. {van Beek}, M.~{Appleby},
  T.~{Raudorf}, J.~{McTiernan}, R.~{Ramaty}, E.~{Schmahl}, R.~{Schwartz},
  S.~{Krucker}, R.~{Abiad}, T.~{Quinn}, P.~{Berg}, M.~{Hashii}, R.~{Sterling},
  R.~{Jackson}, R.~{Pratt}, R.~D. {Campbell}, D.~{Malone}, D.~{Landis}, C.~P.
  {Barrington-Leigh}, S.~{Slassi-Sennou}, C.~{Cork}, D.~{Clark}, D.~{Amato},
  L.~{Orwig}, R.~{Boyle}, I.~S. {Banks}, K.~{Shirey}, A.~K. {Tolbert},
  D.~{Zarro}, F.~{Snow}, K.~{Thomsen}, R.~{Henneck}, A.~{McHedlishvili},
  P.~{Ming}, M.~{Fivian}, J.~{Jordan}, R.~{Wanner}, J.~{Crubb}, J.~{Preble},
  M.~{Matranga}, A.~{Benz}, H.~{Hudson}, R.~C. {Canfield}, G.~D. {Holman},
  C.~{Crannell}, T.~{Kosugi}, A.~G. {Emslie}, N.~{Vilmer}, J.~C. {Brown},
  C.~{Johns-Krull}, M.~{Aschwanden}, T.~{Metcalf}, and A.~{Conway}.
\newblock {The Reuven Ramaty High-Energy Solar Spectroscopic Imager (RHESSI)}.
\newblock {\em \solphys}, 210:\penalty0 3--32, November 2002.

\bibitem[{Mart{\'{\i}}nez Oliveros} et~al.(2012){Mart{\'{\i}}nez Oliveros},
  {Hudson}, {Hurford}, {Krucker}, {Lin}, {Lindsey}, {Couvidat}, {Schou}, and
  {Thompson}]{2012ApJ...753L..26M}
J.-C. {Mart{\'{\i}}nez Oliveros}, H.~S. {Hudson}, G.~J. {Hurford},
  S.~{Krucker}, R.~P. {Lin}, C.~{Lindsey}, S.~{Couvidat}, J.~{Schou}, and W.~T.
  {Thompson}.
\newblock {The Height of a White-light Flare and Its Hard X-Ray Sources}.
\newblock {\em \apjl}, 753:\penalty0 L26, July 2012.

\bibitem[{Mart{\'{\i}}nez-Oliveros} et~al.(2008){Mart{\'{\i}}nez-Oliveros},
  {Moradi}, and {Donea}]{2008SoPh..251..613M}
J.~C. {Mart{\'{\i}}nez-Oliveros}, H.~{Moradi}, and A.-C. {Donea}.
\newblock {Seismic Emissions from a Highly Impulsive M6.7 Solar Flare}.
\newblock {\em \solphys}, 251:\penalty0 613--626, September 2008.

\bibitem[{Matthews} et~al.(2011){Matthews}, {Zharkov}, and
  {Zharkova}]{2011ApJ...739...71M}
S.~A. {Matthews}, S.~{Zharkov}, and V.~V. {Zharkova}.
\newblock {Anatomy of a Solar Flare: Measurements of the 2006 December 14
  X-class Flare with GONG, Hinode, and RHESSI}.
\newblock {\em \apj}, 739:\penalty0 71, October 2011.

\bibitem[{Metcalf} et~al.(2003){Metcalf}, {Alexander}, {Hudson}, and
  {Longcope}]{2003ApJ...595..483M}
T.~R. {Metcalf}, D.~{Alexander}, H.~S. {Hudson}, and D.~W. {Longcope}.
\newblock {TRACE and Yohkoh Observations of a White-Light Flare}.
\newblock {\em \apj}, 595:\penalty0 483--492, September 2003.

\bibitem[{Milligan}(2011)]{2011ApJ...740...70M}
R.~O. {Milligan}.
\newblock {Spatially Resolved Nonthermal Line Broadening during the Impulsive
  Phase of a Solar Flare}.
\newblock {\em \apj}, 740:\penalty0 70, October 2011.

\bibitem[{Mrozek} and {Tomczak}(2004)]{2004A&A...415..377M}
T.~{Mrozek} and M.~{Tomczak}.
\newblock {Solar impulsive soft X-ray brightenings and their connection with
  footpoint hard X-ray emission sources}.
\newblock {\em \aap}, 415:\penalty0 377--389, February 2004.

\bibitem[{Nitta} et~al.(1991){Nitta}, {White}, {Schmahl}, and
  {Kundu}]{1991SoPh..132..125N}
N.~{Nitta}, S.~M. {White}, E.~J. {Schmahl}, and M.~R. {Kundu}.
\newblock {On the reconciliation of simultaneous microwave imaging and hard
  X-ray observations of a solar flare}.
\newblock {\em \solphys}, 132:\penalty0 125--136, March 1991.

\bibitem[{Petrie} and {Sudol}(2010)]{2010ApJ...724.1218P}
G.~J.~D. {Petrie} and J.~J. {Sudol}.
\newblock {Abrupt Longitudinal Magnetic Field Changes in Flaring Active
  Regions}.
\newblock {\em \apj}, 724:\penalty0 1218--1237, December 2010.

\bibitem[{Potts} et~al.(2010){Potts}, {Hudson}, {Fletcher}, and
  {Diver}]{2010ApJ...722.1514P}
H.~{Potts}, H.~{Hudson}, L.~{Fletcher}, and D.~{Diver}.
\newblock {The Optical Depth of White-light Flare Continuum}.
\newblock {\em \apj}, 722:\penalty0 1514--1521, October 2010.

\bibitem[{Ramaty}(1969)]{1969ApJ...158..753R}
R.~{Ramaty}.
\newblock {Gyrosynchrotron Emission and Absorption in a Magnetoactive Plasma}.
\newblock {\em \apj}, 158:\penalty0 753, November 1969.

\bibitem[{Shmeleva} and {Syrovatskii}(1973)]{1973SoPh...33..341S}
O.~P. {Shmeleva} and S.~I. {Syrovatskii}.
\newblock {Distribution of Temperature and Emission Measure in a Steadily
  Heated Solar Atmosphere}.
\newblock {\em \solphys}, 33:\penalty0 341--362, December 1973.

\bibitem[{Sim{\~o}es} and {Costa}(2006)]{2006A&A...453..729S}
P.~J.~A. {Sim{\~o}es} and J.~E.~R. {Costa}.
\newblock {Solar bursts gyrosynchrotron emission from three-dimensional
  sources}.
\newblock {\em \aap}, 453:\penalty0 729--736, July 2006.

\bibitem[{Sim{\~o}es} and {Costa}(2010)]{2010SoPh..266..109S}
P.~J.~A. {Sim{\~o}es} and J.~E.~R. {Costa}.
\newblock {Gyrosynchrotron Emission from Anisotropic Pitch-Angle Distribution
  of Electrons in 3-D Solar Flare Sources}.
\newblock {\em \solphys}, 266:\penalty0 109--121, September 2010.

\bibitem[{Stahli} et~al.(1989){Stahli}, {Gary}, and
  {Hurford}]{1989SoPh..120..351S}
M.~{Stahli}, D.~E. {Gary}, and G.~J. {Hurford}.
\newblock {High-resolution microwave spectra of solar bursts}.
\newblock {\em \solphys}, 120:\penalty0 351--368, September 1989.

\bibitem[{Sudol} and {Harvey}(2005)]{2005ApJ...635..647S}
J.~J. {Sudol} and J.~W. {Harvey}.
\newblock {Longitudinal Magnetic Field Changes Accompanying Solar Flares}.
\newblock {\em \apj}, 635:\penalty0 647--658, December 2005.

\bibitem[{White} et~al.(2011){White}, {Benz}, {Christe}, {F{\'a}rn{\'{\i}}k},
  {Kundu}, {Mann}, {Ning}, {Raulin}, {Silva-V{\'a}lio}, {Saint-Hilaire},
  {Vilmer}, and {Warmuth}]{2011SSRv..159..225W}
S.~M. {White}, A.~O. {Benz}, S.~{Christe}, F.~{F{\'a}rn{\'{\i}}k}, M.~R.
  {Kundu}, G.~{Mann}, Z.~{Ning}, J.-P. {Raulin}, A.~V.~R. {Silva-V{\'a}lio},
  P.~{Saint-Hilaire}, N.~{Vilmer}, and A.~{Warmuth}.
\newblock {The Relationship Between Solar Radio and Hard X-ray Emission}.
\newblock {\em \ssr}, 159:\penalty0 225--261, September 2011.

\bibitem[{White} et~al.(2003){White}, {Krucker}, {Shibasaki}, {Yokoyama},
  {Shimojo}, and {Kundu}]{2003ApJ...595L.111W}
S.~M. {White}, S.~{Krucker}, K.~{Shibasaki}, T.~{Yokoyama}, M.~{Shimojo}, and
  M.~R. {Kundu}.
\newblock {Radio and Hard X-Ray Images of High-Energy Electrons in an X-Class
  Solar Flare}.
\newblock {\em \apjl}, 595:\penalty0 L111--L114, October 2003.

\end{thebibliography}
\end{document}